\documentclass[twocolumn]{aastex701} 

\def \be {\begin{equation}}
\def \ee {\end{equation}}
\def \dd {\mathrm{d}} 
\def \t {\tilde}
\def \p {\partial}
\def \l {\left}
\def \r {\right}

\def \bs {\boldsymbol}

\def \P {\mathcal{P}}
\newcommand{\e}[1]{_{\rm #1}}

\newcommand{\beq}{\begin{equation}}
\newcommand{\eeq}{\end{equation}}
\newcommand{\bea}{\begin{eqnarray}}
\newcommand{\eea}{\end{eqnarray}}

\newcommand{\obs}{_{\rm O}}


\newcommand{\g}{\gamma}

\newcommand{\D}{\DD}
\newcommand{\M}{{\cal M}}
\newcommand{\Y}{Y}

\newcommand{\B}{{\cal B}}

\renewcommand{\th}{\theta}

\renewcommand\S{\mathcal{S}}

\newcommand\n{{\mbox {\boldmath $\nabla$}}}

\newcommand\ees{\end{eqnarray}}
\newcommand\bees{\begin{eqnarray}}

\usepackage{lettrine}
\usepackage{graphicx}
\usepackage{booktabs}
\usepackage{comment}
\usepackage[usenames,dvipsnames]{xcolor}
\usepackage{amsmath}
\usepackage{url}
\usepackage{gensymb}
\usepackage{soul}
\usepackage[utf8]{inputenc}
\usepackage{orcidlink}
\usepackage[normalem]{ulem}
\usepackage[T1]{fontenc}

\def\dd{\mathrm{d}}

\def \obs {\mathrm{obs}}
\def \th {\mathrm{th}}
\def \fid {\mathrm{fid}}
\def \gw {\mathrm{gw}}
\def \gal {\mathrm{g}}

\begin{document}

\title{First measurement of the Hubble constant \texorpdfstring{\\}{ } from 
gravitational wave-galaxy  
cross-correlations}

\author{Isabela Santiago de Matos\orcidlink{0000-0002-2686-2536}}
\email{isabela.matos@port.ac.uk} 

\affiliation{
Institute of Cosmology and Gravitation, University of Portsmouth, United Kingdom
}

\affiliation{ICTP South American Institute for Fundamental Research, São Paulo, Brazil
}

\author{Charles Dalang\orcidlink{0000-0002-7373-6903}}
\email{charles.dalang@ens.fr}

\affiliation{Institut Philippe Meyer, Département de Physique,\\ École Normale Supérieure (ENS), Université PSL, Paris, France  
}

\affiliation{
Institute of Cosmology and Gravitation, University of Portsmouth, United Kingdom
}

\affiliation{Queen Mary University of London, 
United Kingdom}

\author{{Tessa\,Baker}\orcidlink{0000-0001-5470-7616}}
\email{tessa.baker@port.ac.uk}

\affiliation{
Institute of Cosmology and Gravitation, University of Portsmouth, United Kingdom
}

\author{Raul Abramo\orcidlink{0000-0001-8295-7022}}
\email{raulabramo@usp.br}

\affiliation{Instituto de Física, Universidade de São Paulo, 
Brazil}

\author{João Ferri\orcidlink{0009-0005-8123-2528}}
\email{joao.vitor.ferri@usp.br}
\affiliation{Instituto de Física, Universidade de São Paulo, 
Brazil}

\author{Miguel Quartin\orcidlink{0000-0001-5853-6164}}
\email{mquartin@cbpf.br}
\affiliation{Centro Brasileiro de Pesquisas Físicas, 22290-180, Rio de Janeiro, RJ, Brazil}
\affiliation{Observatório do Valongo, Universidade Federal do Rio de Janeiro, 20080-090, Rio de Janeiro, RJ, Brazil}
\affiliation{PPGCosmo, Universidade Federal do Espírito Santo, 29075-910, Vitória, ES, Brazil}

\date{\today }

\begin{abstract} 
We measure for the first time the Hubble constant ($H_0$) from the cross-correlation of galaxies and gravitational waves (GW), by applying the {\it Peak Sirens} method. This method consists of finding the peak of the 3D angular cross-spectrum $C_{\ell}(z,D_L)$ between the galaxy redshifts ($z$) and the GW luminosity distances ($D_L$). Using two GW events from the GWTC-3.0 catalog and the GLADE+ galaxy catalog, we make the first detection  
of the  
cross-correlation peak at $5.9\sigma$
confidence. This signal comes mostly from the best localized event in the catalog, GW190814, which alone provides a $3.4\sigma$ significance. 
Adding also the multimessenger event GW170817, but without using
its known redshift, we find $H_0 = 67^{+18}_{-15}$ 
km s$^{-1}$Mpc$^{-1}$ and the first observational constraint on the GW bias, $b\e{gw} < 4.3$ at 95\% CI.
These measurements set the stage for future novel cosmological constraints with this technique.
\end{abstract}

\section{Introduction}

The past decade of gravitational wave (GW) observations has opened previously uncharted regimes of gravity, cosmology, and astrophysics \citep{Mastrogiovanni:2024mqc, Callister:2024cdx}. For the new field of GW cosmology, the primary focus to date has been constraints on the luminosity distance-redshift relation, which is sensitive to both cosmological parameters and departures from General Relativity (GR) on cosmological scales \cite{Schutz:1986gp, Nissanke:2013fka, Chen:2017rfc, Belgacem:2018lbp}. Such constraints are possible because gravitational waveforms from compact binary coalescences (CBCs) inherently yield the luminosity distance $D_L$ of the source (but not its redshift $z$); this is in contrast with most electromagnetic (EM) sources, whose redshifts, unlike their distances, are generally
straightforward to measure.

Several techniques have been proposed to recover the redshifts of GW sources, with most effort focused on bright siren~\citep{Holz:2005df, Dalal:2006qt, Nissanke:2009kt,  Alfradique:2022tox,Matos:2023jkn} and dark siren methods~\citep{Schutz:1986gp, MacLeod:2007jd, DelPozzo:2011vcw, Mandel:2018mve, Borhanian:2020vyr}. At present, the multimessenger event GW170817 remains the only confirmed bright siren~\citep{LIGOScientific:2017zic,LIGOScientific:2017adf}. Consequently, the dark siren method has risen to become the current workhorse of GW cosmology -- see applications to real data in~\citep{LIGOScientific:2018gmd, DES:2019ccw, DES:2020nay, Gray:2, Quartin:2021dmr, Finke:2021aom, Palmese:2021mjm, Gray:2023wgj, LIGOScientific:2019zcs, LIGOScientific:2021aug, Leyde:2022orh, Chen:2023wpj, Bom:2024afj, LIGOScientific:2025jau}. In particular, the most recent bound on the Hubble constant ($H_0$) obtained by LIGO-Virgo-KAGRA (LVK) \citep{KAGRA:2013rdx, LIGOScientific:2014pky, VIRGO:2014yos, KAGRA:2020tym, Aso:2013eba, Somiya:2011np} from a combined dark+bright siren analysis of GWTC-5.0~\citep{LIGOScientific:2026wfs, Capote:2024rmo, LIGOScientific:2026jgl} 
together with DES Y6 \citep{McMahon:2026nhi} 
is:  
$H_0 = 71.7^{+9.4}_{-7.5}$ km s$^{-1}$Mpc$^{-1}$
at $68.3\%$ confidence interval (CI)~\citep{LIGOScientific:2026uyd}.

In this work we apply to current data a third, different technique for associating redshift information to GW sources: GW-galaxy cross-correlation. Several 
variants of this technique exist \citep{Namikawa:2016edr, Mukherjee:2018ebj, Mukherjee:2020hyn, Diaz:2021pem, Fonseca:2023uay,Balaudo:2023klo,Zazzera:2024agl,Afroz:2024joi, Pedrotti:2025tfg, Sala:2025wwu, Pan:2025iya}
, but in particular we adopt the \textit{Peak Sirens} method put forward in \citep{Ferri:2024amc}, which draws upon on the original work of \citep{Oguri:2016dgk} (see also \citep{Osato:2018mtm, Bera:2020jhx}), an idea akin to the photometric redshift calibration of \citep{Newman:2008mb}.
The Peak Sirens method uses the fact that the spatial correlations between GW events and galaxies, measured in terms of the angular cross-spectrum, peak when the redshifts of galaxies correspond to the luminosity distances of the GW hosts. 
Like other standard siren methods, Peak Sirens bypasses the need for external distance calibrations, 
making it a promising avenue for tackling the Hubble tension \citep{DiValentino:2021izs}; the position of the peak was also shown in~\citep{Ferri:2024amc} to be a robust, model-independent cosmic standard ruler which is  insensitive to the precise shape of the matter power spectrum.

The Peak Sirens method allows constraints on the bias of GW sources ($b_{\rm gw}$), a number quantifying the extent to which GW sources trace the underlying dark matter field. This is a point of difference with current dark sirens approaches, which do not have direct sensitivity to $b_{\rm gw}$.

Another crucial point of difference with the dark sirens method is that the Peak Sirens technique does not rely on assumptions about the mass distribution and merger rate of CBCs. On one hand, features in the mass distribution are a highly informative component of current 
constraints, especially when the galaxy catalog used has low 
completeness over the typical range of the GW events in hand \citep{Ezquiaga:2022zkx, Pierra:2023deu, Li:2024rmi, Farah:2024xub, MaganaHernandez:2025cnu}. On the other hand, this dependence  
introduces a large number of additional parameters that must be constrained alongside the cosmology 
-- up to $25$ in 
\citep{LIGOScientific:2025jau}. Hence, 
Peak Sirens 
is a more direct probe of cosmology with less astrophysical entanglement.

\textit{In this work} we perform the first application of the Peak Sirens method to data using GW190814 and GW170817 (though the redshift of its host galaxy  
is not used) 
and the K-band galaxies of the GLADE+ catalog (see however \footnote{ We note that an attempt to detect the GW-galaxy cross-correlation signal using similar data sets was presented in \citep{Mukherjee:2022afz}, but returning a null result. This work has various differences in methodology compared to ours. In particular, those authors have projected the angular power spectra in the distance dimension, $C_\ell(z,D) \to C_\ell(z)$, wiping out the correlations between different bins of redshift and distance, which is essential to infer the very $D_L(z)$ relation that we are trying to recover.}
). As very few events lie within the depth of the GLADE+
catalog, GW170817 is included here 
to help demonstrate the viability of the method, even if tighter $H_0$ constraints might be obtained instead with its use as a bright siren. Since we have good galaxy coverage for only a limited number of GWs,
we focus on the dominant parameter sensitivities: $H_0$ and $b_{\rm gw}$.

\section{Methodology}

\noindent{\bf 3D angular cross-correlations.} 
A key assumption of this work 
is that GW sources are located inside galaxies, 
such that their positions 
are correlated. 
To study these correlations, we build pixelated galaxy and GW source sky maps using
\texttt{HEALPix}~\citep{Zonca:2019vzt, Gorski:2004by}. 
For galaxies, we decompose the radial direction in redshift bins $z_i$, which, together with the \texttt{HEALPix} pixels, fully defines voxels. 
We count the number of galaxies $N_{\gal}(z_i, \bs{\hat{n}})$ in each voxel denoted by the 
direction $\bs{\hat{n}}$ and redshift bin center $z_i$. 
We integrate the  
$z$-PDF assuming Gaussian uncertainties over each voxel. 
This map encodes the PDF of finding galaxies in each voxel. For the GW data, we decompose instead the radial direction in $D_L$ 
bins indicated by their center $D_{L j}$. We integrate the observed 
3D location posterior, marginalized over other binary parameters like masses and spins, to count the fractional GW contribution to each voxel. This defines the GW map $N_{\gw}(D_{Lj}, \bs{\hat{n}})$, which is effectively the sum of the 
PDFs of the GW sources.

With these two maps, one can compute their angular cross-correlation:
\begin{align}
    \textstyle
    C_{\ell}(z_i, D_{Lj}) = \frac{1}{2\ell +1}\sum_{m = -\ell}^{\ell} a_{\ell m}^{\gal}(z_i) \, a^{\gw *}_{\ell m}(D_{Lj}) \,, \label{eq:CL}
\end{align}
where $a_{\ell m}^{\gal}(z_i)$ and $a_{\ell m}^{\gw}(D_{Lj})$ are the coefficients of their respective spherical harmonic decompositions:
\begin{equation}
    \textstyle
    N_{X}(r_{Xi},{\bs{\hat{n}}}) = \sum_{\ell m} a^{i}_{\ell m}(r_{Xi}) Y_{\ell m}({\bs{\hat{n}}}) \, ,
\end{equation}
where in turn $X=\{\gal,\gw\}$ for galaxies and GW sources, respectively, $r_{X}$ is either $z$ or $D_L$, and $Y_{\ell m}(\bs{\hat{n}})$ are the spherical harmonics. 
The resulting 
coefficients $C_{\ell}(z_i,D_{Lj})$ correspond to the 3D angular power spectrum after removing shot noise.

We do not use auto-correlations in our data vector, since our focus is detecting the peak of the cross-spectrum and inferring the background cosmology from this new observable. The clustering (auto-correlation) of galaxies is a key cosmological probe of the linear perturbations; here we will use only its amplitude (and not the shape) to eliminate a global degeneracy in 
the power spectra between the GW and galaxy biases and the clustering amplitude $\sigma_8$, since the cross-correlation amplitude depends on the product $b_{\gal}b_{\gw}\sigma_8^2$.  
We find that the 
GW auto-correlation with current data is not yet informative due to its high shot noise, as expected~\citep{Zazzera:2024agl,Namikawa:2015prh}. 
We highlight a common misconception: that many GW events are needed to detect a GW-galaxy cross-correlation. This intuition is based on auto-correlation measurements -- see  Appendix B for discussion. 
We model our data vector, 
defined in Eq.\,\eqref{eq:CL}, with a Gaussian likelihood, 
\begin{equation}
    \textstyle
    \log(\mathcal{L}) = -\frac{1}{2}\sum_{\ell ij, \ell' i'j'} \Delta C_{\ell}^{ij} \, [{\rm Cov^{-1}}]_{\ell ij, \ell' i'j'} \, \Delta C_{\ell'}^{i'j'} \; , \label{loglike}
\end{equation}
where
$\!\Delta C_{\ell}^{ij} \!\!\equiv\! C_{\ell}^{\obs}(z_i, D_{Lj}) - C_{\ell}^{\th}(z_i, D_{Lj})$
denotes the difference between observed and theoretical cross-spectra. 

\vspace{6pt}

\noindent{\bf Theoretical modelling.} 
We use simulations to model both  
$C_\ell^{\th}(z_i, D_{Lj},\bs{\theta})$ in an arbitrary cosmology denoted by $\bs{\theta}$, and the covariance matrix in Eq.\,\eqref{loglike}. Pairs of GW and galaxy catalogs are built in a single fiducial model $\bs{\theta}^{\fid}$, with GW sources injected randomly into galaxies, as we describe in the appendix. 
Averaging the $C_\ell$'s  
over simulations allows us to obtain an accurate value for $C_\ell^{\fid}(z_i, D_{Lj})$. For a fixed redshift bin $z_i$, the cross-spectrum peaks 
at $D_L(z_i, \bs{\theta}^{\fid})$ due to the rapid decline of the two-point correlation function of the matter field with comoving distance. 

When changing continuously the cosmological parameters (here just $H_0$), we account for the change in 
this single feature, namely the position of the sharp peak in the $(z,D_L)$ plane, while neglecting the cosmology dependence of the cross-spectrum shape, 
which is a sub-leading effect 
\citep{Ferri:2024amc}. The position of the peak is adapted by rescaling the redshift axis of $C_\ell^{\fid}(z_i, D_{Lj})$  as in \citep{Ferri:2024amc}, which can be done because the mapping between $z$ and $D_L$ 
is one-to-one for a given cosmology
~\footnote{There is a slight difference here from~\citep{Ferri:2024amc},
where instead distance bins are rescaled, i.e. $C_{\ell}^{\th}(z_{i},D_{Lj}, \bs{\theta}) = C_{\ell}^{\fid}(z_i,z(D_{Lj},\bs{\theta}))$.
In this
work, we rescale the galaxy redshift bins according to the new cosmology. This is more robust against the density profiles of each GW event since there are very few GW events versus an abundant numbers of galaxies.}. 
This method allows us to model the theoretical cross-spectrum for different cosmologies by doing simulations in only one fiducial cosmology, which we choose to be the CMB primary+lensing result of \citep{Planck:2018vyg} (in particular, $H_0 = 67.36$ km s$^{-1}$Mpc$^{-1}$ and $\Omega\e{m} = 0.3153$).

\vspace{6pt}

\noindent{\bf On the gravitational wave bias.} The linear GW bias modulates 
the amplitude of the cross-correlations (see Eq.\,(2.16) of \citep{Ferri:2024amc}), and as long as it varies slowly with redshift, it does not affect the position of the peak significantly. We parametrize it as $b\e{gw}(1+z)$. The free parameter $b\e{gw}$ should be interpreted as the mean GW bias of the particular sample of GW sources, for which bias stochasticity can be significant~\citep{Dekel:1998eq,Paranjape:2017zpc}.

For Gaussian random fields, the covariance matrix of the angular power spectra depends negligibly on the bias of the lower-density field in the limit of low number density ($\bar{N}_{\gw} \ll \bar{N}_{\gal}$). This is due to high shot noise of the auto-correlations. Taking the Gaussian case as inspiration, we will assume that  
$b\e{gw}$ enters in the likelihood only as a free factor multiplying $C_\ell^{\rm{th}}$, which is estimated from simulations with a fixed GW bias of $1$.

\section{Data}

\noindent{\bf GLADE+ maps.} We use K-band data of the GLADE+ catalog \citep{Dalya:2021ewn} to construct a galaxy number count map, $N_{\mathrm{g}}(z,\bs{\hat{n}})$. This catalog has been widely used for GW cosmology with dark sirens~\citep{LIGOScientific:2021aug,Gray:2023wgj,Mastrogiovanni:2023emh,Chen:2023wpj,LIGOScientific:2025jau} which allows for direct comparison with these results. We use 
$n\e{side} = 32$ to split the sky into equal-area pixels of $\sim 4$ deg$^2$. We fix the voxel depth to $\Delta z = 0.01$ in most of the redshift range; for $z<0.02$, where we have precise galaxy redshifts, we use bins 3$\times$ smaller instead, which is crucial to locate the cross-correlation peak of the nearest GW sources. After that, our largest bin size is of the order of the photometric redshift uncertainties that we assume to be Gaussian for simplicity. 
It was shown in \citep{Turski:2023lxq} that this assumption does not affect significantly current measurements of $H_0$.  
While this simplistic choice matches current LVK dark sirens work \citep{LIGOScientific:2025jau}, further refinements might be required when the precision on $H_0$ increases.  
Furthermore, we make a conservative cut removing all galaxies whose photo-z Gaussian PDF would have more than 0.15\% support below $z=0$. 

The observed number count map $N\e{g}(z, \bs{\hat{n}})$ relates to the a priori unknown \textit{complete} galaxy map  
via the completeness $\hat{f}_\S(z,\bs{\hat{n}}) \in [0,1]$, which we estimate from the magnitude threshold $m\e{th}(\bs{\hat{n}})$ of each pixel, 
following~\citep{Gray:2021sew}: 
\begin{align}
    \hat{f}_\S(\bs{\hat{n}},z) = \textstyle \int_{L(m\e{th}(\bs{\hat{n}}))}^{L\e{max}} \dd L\, \phi(L)  \Big/{\textstyle \int_{L\e{min}}^{L\e{max}} \dd L\, \phi(L) }\,, \label{eq:completeness}
\end{align}
where the luminosity function $\phi(L)$ obeys a Schechter function~\citep{Schechter:1976iz}.
For the GLADE+ K-band, we use $L_*$, $L\e{min}$ and $L\e{max}$ converted from the absolute magnitudes $M_* = -23.39$, $M\e{min} = -27$ and $M\e{max}=-19$, with $\alpha=-1.09$~\citep{Kochanek:2000im}. 
We use 
the $k$-correction  estimate \citep{Hogg:2002yh} from the python package 
\texttt{kcorrect} and additional information from the J-, H- and B-bands~\citep{Blanton:2006kt}. 
The luminosity is then calculated via $L(M) = L_0 \cdot 10^{-2 M /5}$ with $L_0 = 3.0128 \cdot 10^{28}$ W. 
We apply a galactic latitude mask $|b|>20^\circ$ to get rid of star contamination close to the Milky Way. Finally, we use the publicly available code \texttt{Namaster}~\citep{Alonso:2018jzx} to estimate the auto and cross-correlations from the masked data, which has after all cuts approximately $2.8\times 10^5$ galaxies.

\vspace{6pt}

\noindent\textbf{{GWTC-3.0.\,}}
Most GWs in GWTC-3.0 \citep{KAGRA:2021vkt, Buikema2020} lie outside the GLADE+ redshift range, when 
computing $z(D_L)$ with some value of $H_0$ in the prior range of $[20,120]$. This results in three candidates for our analysis: GW190425, GW190814, and GW170817. The first is likely a binary neutron star (BNS) merger, observed only by LIGO Livingston, 
resulting in poor localization. This event was thus excluded from our analysis. For the same reason, we do not consider events from GWTC-4.0 (detections from the O4a observing run), as these were generally poorly localised due to the lack of Virgo observations. The second event, GW190814, is the merger of a black-hole and a low mass compact object and it is the best-localized event in the catalog, with area of $\Omega_{90} = 18.5\deg^2$ at 90\% CI. 
We have additionally GW170817,  the single bright siren observed so far, whose host galaxy NGC 4993 is contained in GLADE+. We therefore focus the analysis on these last two events. 

\begin{figure*}[t]
    \centering
    \includegraphics[width=0.75\linewidth]{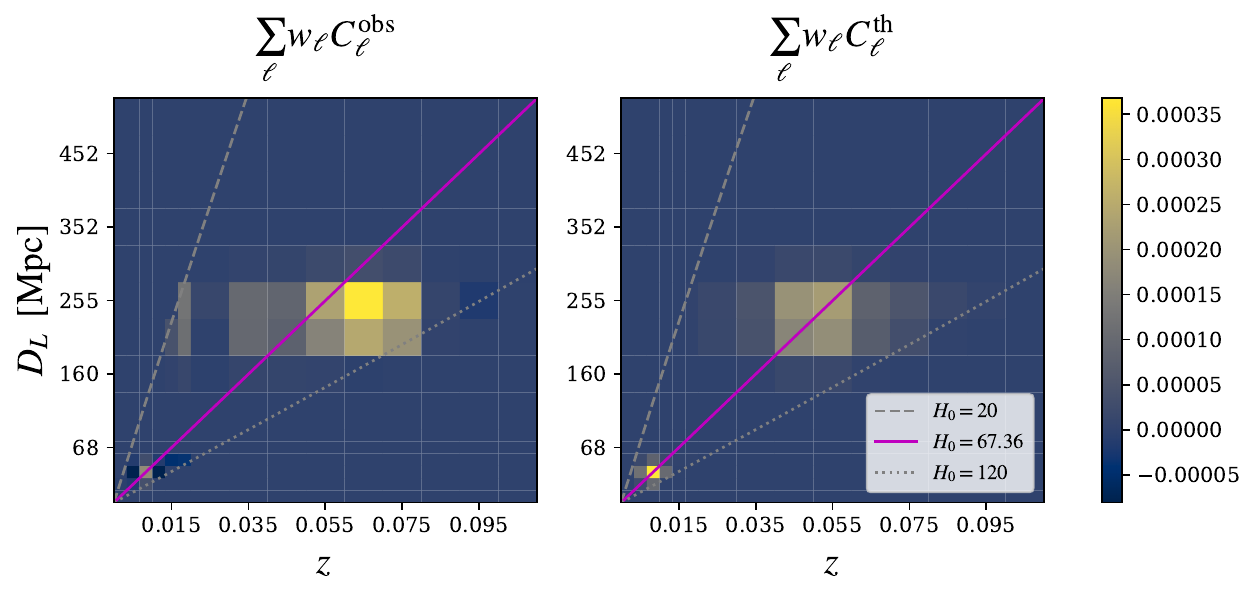}
    \vspace{-10pt}
    \caption{ Angular cross-correlations between GLADE+ and GW190814, GW170817. \emph{Left:} Sum over multipoles $\ell \in [14,69]$ of the $C_\ell$s as a function of galaxy $z$ and GW $D_L$. 
    Regions corresponding to $H_0$ outside the prior range $[20,120]$ \textrm{km s$^{-1}$ Mpc$^{-1}$} (grey dashed and dotted curves) are set to zero for visualization purposes. The magenta curve illustrates how the Planck cosmology compares with the data. 
    \emph{Right:}
    Same but obtained from an average over our set of simulations in the fiducial cosmology, with $b_{\gw} = 2$. 
    The bright regions correspond to the peaks in the cross-correlation signal generated by events GW170817 (bottom left corner) and GW190814 (central region)  and reveal the Hubble diagram.}
    \label{fig:cl_data}
\end{figure*}

For GW190814, the GW map is built from the GW posterior obtained with the waveform IMRPhenomXPHM \citep{LIGOScientific:2021usb}. For GW170817, we use the GW result with 
PhenomPNRT with low spin prior, having fixed the sky location of the host galaxy~\citep{LIGOScientific:2018hze}. 
We do not use, however, any information about the measured redshift of the bright siren's host galaxy.  
For convenience, we take $D_L$ 
bins that correspond, in the fiducial cosmology,  
to the redshift binning 
used for galaxies.
We also bin the multipoles, with bin size $\Delta \ell = 5$, which minimizes correlations 
introduced by the mask. We exclude the first two small-$\ell$ bins which are uninformative due to cosmic variance and the galactic plane mask. 
We further exclude scales smaller than the resolution of  GW190814 ($\ell > 180\deg/\sqrt{\Omega_{90}/\pi}$), such that our bin centers are $\ell \in \{14, 19, 24, ..., 69\}$. 
We show on the LHS of Fig.\,\ref{fig:cl_data} a summary of the data as a weighted sum of $C_\ell$'s. The weights $w_{\ell} = (2\ell+1)P_{\ell}(\cos(1.8\degree))$, where $P_{\ell}$ are the Legendre polynomials, correspond to a beam with the pixel angular resolution. GW170817 and GW190814 are located, respectively, around the third $D_L$ bin at $\simeq$ 40 Mpc, and at 255 Mpc. We can identify the corresponding two peaks at redshifts $\simeq$ 0.008 and 
$z\simeq 0.065$, the same structure one can observe on the RHS plot, which shows the theoretical prediction from almost 6000 simulations, that are described in the appendix.

\section{Results}

\begin{figure}[t]
    \centering
    \includegraphics[width=.9\linewidth]{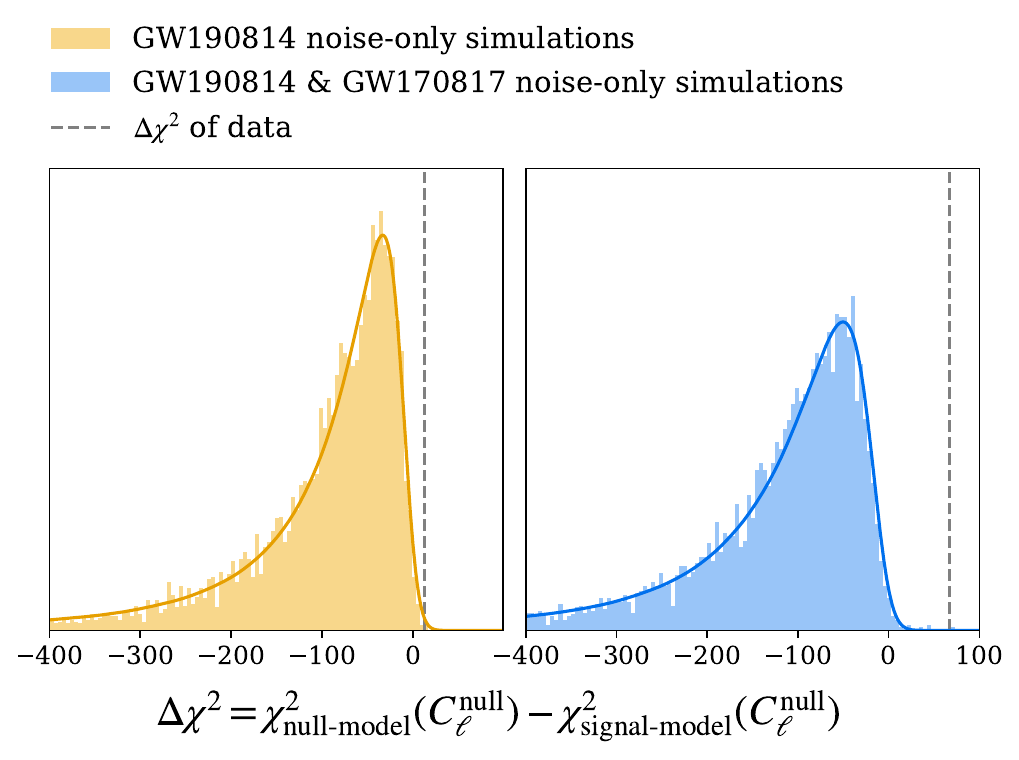}
    \caption{$\Delta \chi^2$ between the null hypothesis (noise-only) and the model with a cross-correlation signal, for correlations with GW190814 only (left) and additionally GW170817 (right). 
    The data (grey dashed lines) considerably prefers, with its large $\Delta \chi^2$, the signal-model against the noise-model.
    The solid lines are derived from the Fréchet distribution fits for the individual $\chi^2$ histograms.
    }
    \label{fig:pvalue}
\end{figure}

\noindent{\bf Significance of detection.} To determine the significance of detection of cross-correlations, we take galaxy and GW maps from different simulations, such that they are uncorrelated, and compute their cross-spectrum 
$C_{\ell}^{\rm null}$, as well as its average and sample covariance. We build the $\Delta \chi^2$ distribution of these noise-only simulations to quantify how often a simulation under the null hypothesis (no cross-correlations) appears closer to the average of correlated maps than to the average of uncorrelated maps. We plot this $\Delta \chi^2$ distribution in Fig.\,\ref{fig:pvalue} and compare with the $\Delta \chi^2$ of the data.
We find that the data favors
the presence of cross-correlations over noise. Fitting Fréchet distributions to each individual $\chi^2$ of the simulations, we estimate the p-value to be $6.7 \times 10^{-4}$
for cross-correlations with GW190814 only, equivalent to a two-sided Gaussian significance of $3.4\sigma$
Adding GW170817 decreases the p-value  to 1.9 $\times 10^{-9}$,
which is equivalent to a significance of
$5.9\sigma$. 
Notice that this significance is achieved over all $(z, D_L)$ bin pairs and all scales in our range together.

\vspace{6pt}

\begin{figure}[t]
    \centering
    \includegraphics[width=.9\linewidth]{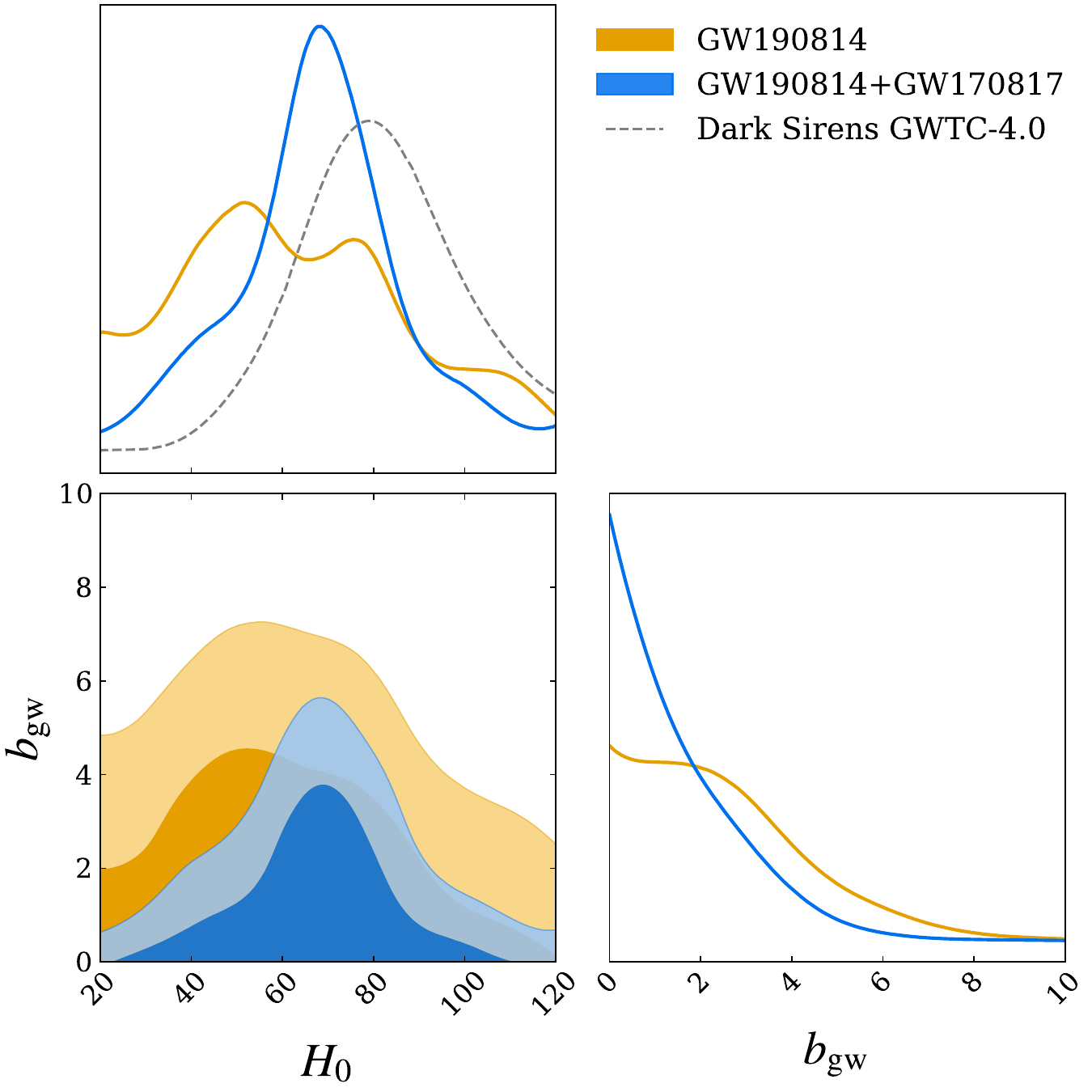}
    \caption{68\% and 95\% CI for the cross-correlations of GLADE+ with only GW190814 (orange) and with both  GW190814 and the BNS GW170817 as a dark siren (blue). We also show the $H_0$ posterior from GWTC-4.0 dark sirens analysis (grey, dashed) for reference \citep{LIGOScientific:2025jau}.}
    \label{fig:mcmc_contours}
\end{figure}

\noindent{\bf Hubble constant and GW bias. 
}We run MCMC chains to infer the Hubble constant and GW bias, as shown in Figure~\ref{fig:mcmc_contours}. We assume the flat priors $H_0 \in [20,120]$ \textrm{km s$^{-1}$ Mpc$^{-1}$} and $b_{\gw} \in [0,10]$. 
In order to measure $b_{\gw}$ alone, we estimate $b_{\gal}$ by comparing the galaxy auto-spectrum of the data with the one from the average over simulations, after removing shot noise. From the amplitude of their ratio we obtain $b_{\gal}(\sigma_8/\sigma_8^{\fid}) = 0.85 \pm 0.13$. We then fix $\sigma_8$ to the very precise CMB measurement, $\sigma_8 = 0.8111 \pm 0.0060$ \citep{Planck:2018vyg}. 
With GW190814 only, we find the following marginalized constraints (highest-density intervals):
\begin{align}
    H_0 &= 53^{+31}_{-18}
    \;\; \textrm{km s$^{-1}$ Mpc$^{-1}$}\;\; \textrm{(at 68\% CI)}\,, \\
    b_{\gw} &<6.2 \;\; \text{(at 95\% CI)}\,. 
\end{align}
Adding GW170817 without its $z$ information leads to
\begin{align}
    H_0 &= 67^{+18}_{-15} 
    \;\; \textrm{km s$^{-1}$ Mpc$^{-1}$}\;\; \textrm{(at 68\% CI)}\,, \\
    b_{\gw} &< 4.3 \;\; \textrm{(at 95\% CI)}\,. 
\end{align}
We verified from the $\chi^2$ distribution of the simulations that our best fit in each case is a good fit to the data. 

In contrast, adding GW170817 as a bright siren to the Peak Siren measurement with GW190814, one obtains $H_0 = (72 \pm 11)$ km s$^{-1}$Mpc$^{-1}$. 
If we instead combine the two Peak sirens with the spectral siren \textsc{MLTP} result from~\citep{LIGOScientific:2025jau}, which uses information from the black hole mass distribution, we obtain $H_0 = 68^{+16}_{-8}$
km s$^{-1}$Mpc$^{-1}$ as a first approximation, neglecting correlations between the two analysis.  Peculiar velocity effects were not included in this work; while for galaxies they are expected to be negligible with those bin sizes, the aberration in the distances could generate a shift in $H_0$ which is smaller than our final uncertainties (see e.g. \citep{Nicolaou:2019cip}). 

We can also constrain 
the GW bias relative to galaxies instead of the dark matter (which is the normal implication of $b_{\gw}$). Since this is roughly the ratio between the cross-spectrum and the galaxy auto-spectrum, it does not depend on $\sigma_8$. We find $b_{\gw}/b_{\gal} < 8.1$ with GW190814 and $b_{\gw}/b_{\gal} < 5.7$ with both GW190814 and GW170817. This quantity is relative to the galaxy population we are analyzing, in this case, galaxies with luminosities in the K-band, which are generally regarded as  tracers of stellar mass. 

Finally, we remark that the two parameters $H_0$ and $b_{\gw}$ have an almost transversal effect on the cross-correlations; while the first sets the peak position, the feature most precisely measured, the second governs its amplitude. For a fixed distance, however, the redshift of the peak also implies the number of observed galaxies at the peak, and the cross-correlation amplitude scales with $\bar{N}_{\gal}(z)$, that being the source of correlations between the two parameters. Furthermore, the triangular shape of the contours in Fig.\,\ref{fig:mcmc_contours} reflect the scaling of the signal-to-noise ratio with the GW bias, whose value directly impact the precision in $H_0$. In addition to this degeneracy, the need to model how the covariance depends on $b_{\gw}$ as it approaches zero could be the reason why we can only place an upper bound on this parameter, despite making a significant detection of cross-correlations.

\vspace{6pt}

\section{Discussion}

In this work, we have presented the first detection of the GW-galaxy cross-correlation signal. Even with a restricted set of only two GW events, we have shown that it is possible to measure this signal: this is achievable due to the large number of GW-galaxy pairs that can be constructed in the data. For this detection to be attainable, it is also crucial to use sufficiently narrow tomographic bins, and to accurately model the covariance of the data (which we have done via constrained simulations.)
From the peak location of the cross-correlation signal we obtain a measurement of 
$H_0 = 67^{+18}_{-15}$ km s$^{-1}$ Mpc$^{-1}$. While that is not yet competitive with state-of-the-art measurements from the CMB and standard candles, it is comparable (and compatible)
with other dark siren measurements of $H_0$~\citep{DESI:2023fij,Vasylyev:2020hgb,LIGOScientific:2021aug,Bom:2024afj,Alfradique:2023giv,DES:2020nay,Palmese:2021mjm,Alfradique:2023giv,Finke:2021aom}, revealing the strength of the Peak Sirens method.

Assuming those events can be described by a single bias parameter at this point, we report its first observation constraint $b\e{gw} < 4.3$. 
Although this number is subject to a high degree of stochasticity due to our small GW sample, it must be consistent with the `true' GW bias, and converge to in the limit of many GW sources. GW bias will play a key role in the future as it relates to the formation mechanisms of compact objects~\citep{Scelfo:2018sny}, and many forecasts were made for 3G constraints~\citep{Libanore:2020fim, Libanore:2021jqv, Zazzera:2024agl, Zazzera:2025ord, Pedrotti:2025tfg, ET:2025xjr}; our bound here marks the start of that process. 

Current dark sirens results are primarily informed by the presence of features in the mass spectrum of compact objects. 
In this work, when combined with GW observations, the information on $H_0$ comes only from the galaxy catalog. 
Further precision improvements can be achieved by properly combining Peak Sirens with spectral sirens \citep{Cheng:2026atn}.  
Furthermore, in the dark sirens technique, a redshift prior for GW events (constructed from a galaxy catalog) is applied along individual lines of sight within the GW sky location 
\citep{Gray:1}. As such, the clustering of galaxies enters the analysis along radial lines of sight only, while the Peak Sirens approach explicitly incorporates the 3D clustering. A comprehensive study of the relation and covariance between these methods is underway.

In the simulations used in this work (see Appendix), GW host galaxies were not assumed to be necessarily in the galaxy catalog generated after applying realistic selection effects. Therefore, the modeling of the cross-correlation signal  did not rely on the exact coincidence of the positions object-by-object, in line with the predictions of \citep{Bera:2020jhx, Ferri:2024amc}. At the same time, in those simulations we do assume that all galaxies are in principle equally likely to host a CBC. Weighting galaxies according to their chance of hosting CBCs is a component of the dark siren method that relies on modeling this probability as a function of galaxy properties such as luminosity or stellar mass \citep{Gray:2, Palfi:2025wua, Li:2025hrh, Hanselman:2024hqy}. On one hand, this introduces more entanglement with astrophysical modeling and, on the other hand, it can enhance correlations. 
Here, instead, the link with galaxy properties is encoded in the 
GW bias \citep{Dehghani:2024wsh,Hosseini:2025plz}. An alternative approach that includes galaxy weighting would require simulations where not only positions but also those properties are predicted.

In the near future, 
we intend to use a deeper galaxy catalog (e.g.~GAIA, DESI Legacy), which has more  
support at the  distances relevant to GWTC-4.0, since the shallowness of GLADE+ was the main limitation of the present work. Improvements are expected by including further well-localized GW events from GWTC-3.0 
 and eventually from the latter stages of the LVK O4 run.

As we accumulate new GW data, the precision on $H_0$ depends crucially on the sky position errors of GW sources. At fixed resolution, since $\bar{N}_{\gw}\ll \bar{N_{\gal}}$
, the signal-to-noise ratio of angular cross-spectrum in the Gaussian field approximation scales with $\sqrt{\bar{N}_{\gw}}$. With 100 events with skymaps resembling GW190814, one expects a $\sim 4\%$ 
measurement of the $H_0$ using this method.

Finally, Peak Sirens 
is ideal for tests of gravity theories in which distances measured from GWs differ from the standard luminosity distance~\citep{Lombriser:2015sxa, Saltas:2014dha, Amendola:2017ovw, Belgacem:2018lbp, Nishizawa:2017nef, Baker:2020apq, Leyde:2022orh,Matos:2023jkn, Chen:2023wpj, Colangeli:2025bnb, Mukherjee:2020mha}. We plan to use this method to measure the dark energy equation of state and the friction induced in tensor modes. This work represents the first stone of a castle of many opportunities, which are left for future work.

\vspace{6pt}

\begin{acknowledgements}

We would like to thank Fabien Lacasa, Louis Legrand, Konstantin Leyde, 
Krishna Naidoo, Riccardo Sturani, Nicolas Tessore and Michael Williams for useful discussions and insights. We also thank Nicola Borghi, David Alonso and Maciej Bilicki for comments on the draft.
I.S.M.\,, C.D.\,and T.B.\, are supported by ERC Starting Grant SHADE (grant no.\,StG 949572). 
I.S.M.\,also acknowledges funding from FAPESP. T.B.\,is further supported by a Royal Society University Research Fellowship (grant no.URF\textbackslash R\textbackslash 231006).
R.A. is supported by FAPESP and CNPq. J. F. would like to thank CNPq for financial support. M.Q. is supported by the Brazilian research agencies FAPERJ project E-26/201.237/2022, CNPq and CAPES. This material is based upon work supported by NSF's LIGO Laboratory which is a major facility fully funded by the National Science Foundation.

Supporting research data are available on reasonable request from the corresponding authors. For the purpose of open access, the author(s) have applied a Creative Commons Attribution (CC-BY) license to any Author Accepted Manuscript version arising. The data behind figures can be found in~\dataset[doi:10.5281/zenodo.21908857]{https://zenodo.org/records/21908857}.

\end{acknowledgements}

\appendix

\twocolumngrid

\section*{Appendix A: Simulations} 
\label{sec:simulations}

\vspace{6pt}

\noindent{\bf Galaxy simulations.} We use the code GLASS~\citep{Tessore:2023zyk} to perform galaxy simulations. We give as an inputs: the 3D matter power spectrum obtained from CAMB~\citep{camb:2011} in the fiducial cosmology,  
the galaxy bias $b_{\gal} = 1 + z$ 
and the total number of galaxies 
in each redshift bin. 
The code builds a lognormal field on the light-cone that has these properties.
Hence, we simulate a complete version of the GLADE+ catalogs, meaning that we generate $N\e{c}(z)$ galaxies. This profile is slightly deformed with respect to the estimate from the completeness fraction, due to photo-z smearing, such that the final redshift distribution recovers the observed one. 
We use this completed galaxy map to draw the GW host galaxies. The final map, which will mimic only the observed galaxies will therefore not necessarily contain the GW host galaxies. 

We then build an observed version $N\e{g}(z,\bs{\hat{n}})$ of this catalog, which is masked around the Milky Way, like the data. Since no luminosity function is assigned to galaxies in our simulations, to model a magnitude-limited survey, we exclude galaxies proportionally to the completeness map in each pixel and redshift bin. Note that this step assumes that the unobserved galaxies have the same clustering statistics as the observed ones.
We then draw a number count from a Poisson distribution centered around the number of remaining galaxies in each voxel. Finally, based on the observed fraction of galaxies with photometric redshifts, we sample an updated redshift from a Gaussian and integrate the PDF of these galaxies over each voxel to match what is done with the data, while keeping the fraction of spectroscopic galaxies point-like.

\vspace{6pt}

\noindent{\bf Gravitational wave simulations.} 
We perform constrained simulations
which enforce  
the statistical properties of the true data sets, that is, we always simulate the ``same" GW events, but as if they were happening in different galaxies.
Due to the small number of GW events,
we fix the angular and radial resolutions of the simulated GW events to be the same as those, but sample the radial and sky positions randomly as follows. 
First, for each GW  
in each simulation, we draw from the real data marginalized $D_L$  
posterior the true value of $D_L$ for that simulation. 
We then convert the latter to a redshift using the fiducial cosmology. 
The sky position is found by randomly picking a host galaxy from the corresponding redshift bin $z_i$.
Given this ``true'' position of the GW event, its 3D posterior is placed such that the peak of the distribution 
coincides with the sky position of the host galaxy. 
Finally, we draw from this distribution to obtain the ``measured'' position of the GW event, and re-center the distribution around that. 
We assume a simple scenario where all galaxies are equally likely to host a GW, and therefore, $b_{\gw}/b_{\gal} = 1$ is the fiducial value in the simulations; folding in correlations between galaxy properties and the probability of hosting is left for future work.

Since the two events considered were observed well outside the masked region, we exclude simulations where they fall more than 5\% volume-wise 
in the mask, in order to avoid an over-estimation of the covariance matrix. The anisotropic antenna patterns of the detectors are not being reproduced in our simulations, for simplicity, as with only a small number of events, anisotropies in the GW distribution are dominated by shot noise.
We generated almost 6000 simulations, 
from which the covariance in Eq.\,\eqref{loglike} 
could be accurately estimated. 

We have also generated simulations where galaxies (and consequently also GWs) are drawn uniformly in the sky. The cross-spectrum of those maps were used to estimate the shot noise contribution to our cross-correlations, which was 
negligible and was removed from our data vector.

\section*{Appendix B: Common misconceptions} 
\label{sec:misconceptions}

\vspace{6pt}

\noindent{\bf Clustering of GWs vs cross-correlations.} A common misconception about our measurement comes from the intuition built from auto-correlations (i.e. clustering). In this work we are not measuring clustering of GWs, as this is impossible with such a sparse sample due to high shot noise. We are nonetheless assessing the clustering of structures via probing the galaxy clustering at the GW position -- which is in fact a region, defined by the probability of the true source position given the marginalized posteriors from the GW strain data. Moreover, the statistics of the GW-galaxy cross-spectrum does not necessarily follow a Gaussian statistic, which is a common approximation in LSS analysis {\it for many point-like sources}. GWs are not being drawn with Poissonian distribution around the dark matter density -- the total number of objects is fixed (to two) in all simulations and that changes the variance of this field as well. Also, GW true positions are drawn from galaxy positions and not directly from dark matter clumps. However, due to the galaxy selection function part of those hosts are removed, and due to the GW position errors, the exact coincidence of GW positions with galaxies is smeared out, reducing the shot noise contribution to the cross-spectrum. The resulting non-trivial statistics is therefore captured in the simulations and embedded in our {\it sample covariance} (a point of difference with our forecast paper \citep{Ferri:2024amc}).

\vspace{6pt}

\noindent{\bf Peak, Dark \& Spectral sirens.} Spectral sirens is the method for GW-redshift inference relying on assuming a source-frame mass distribution for compact objects, and looking how features in this distribution shift at the observer. A further prior on redshift is modeled as growing with the product of comoving volume and  merger rate. Despite the terminology, Dark Sirens contain Spectral sirens as a special case; here the redshift prior is taken instead from a galaxy catalog, stacking all galaxies along the lines of sight contained in the $90\%$ probability area of location of the GW source. Peak Sirens is a fundamentally different method that is based on 2-point statistics and not this unified Bayesian framework. The redshifts are inferred by maximizing correlations between the probability density field of the GW 3D position  with the galaxy distribution, specially on the large scales, and down to the angular scale of the GW sky resolution; therefore, not using galaxies along the line of sight at the GW position. See \citep{Cheng:2026atn} for a recent discussion on the various methods.

\vspace{6pt}

\noindent{\bf Completeness vs selection function.} For the Dark Sirens method, it is important to model the fraction of unobserved galaxies in a given luminosity band, in each pixel and redshift -- called completeness. Since Peak Sirens is based on 2-point statistics of anisotropies, only the relative difference of sensitivity between pixels is important, as for standard LSS analysis pipelines. Therefore, estimating the completeness by fitting a luminosity function is sufficient for both methods, but for Peak Sirens the denominator of Eq.\,\eqref{eq:completeness} is irrelevant as it is a global factor for all pixels, and therefore, the precise value of $L\e{min}$, is irrelevant.

\nocite{Mukherjee:2022afz}

\bibliographystyle{aasjournalv7}
\bibliography{references}

\end{document}